# Scaling Behavior of Quantum Nanosystems: Emergence of Quasi-particles, Collective Modes, and Mixed Exchange Symmetry States

Zeina Shreif and Peter Ortoleva

Department of Chemistry
Indiana University
Bloomington, IN 47405
Contact: *ortoleva@indiana.edu*

**Abstract**

Quantum nanosystems such as graphene nanoribbons or superconducting nanoparticles are studied via a multiscale approach. Long space-time dynamics is derived using a perturbation expansion in the ratio $\varepsilon$ of the nearest-neighbor distance to a nanometer-scale characteristic length, and a theorem on the equivalence of long-time averages and expectation values. This dynamics is shown to satisfy a coarse-grained wave equation (CGWE) which takes a Schrö dinger-like form with modified masses and interactions. The scaling of space and time is determined by the orders of magnitude of various contributions to the $N-$body potential. If the spatial scale of the coarse-graining is too large, the CGWE would imply an unbounded growth of gradients; if it is too short, the system's size would display uncontrolled growth inappropriate for the bound states of interest, i.e., collective motion or migration within a stable nano-assembly. The balance of these two extremes removes arbitrariness in the choice of the scaling of space-time. Since the long-scale dynamics of each fermion involves its interaction with many others, we hypothesize that the solutions of the CGWE have mean-field character to good approximation, i.e., can be factorized into single-particle functions. This leads to a Coarse-grained Mean-field (CGMF) approximation that is distinct in character from traditional Hartree-Fock theory. A variational principle is used to derive equations for the single-particle functions. This theme is developed and used to derive an equation for low-lying disturbances from the ground state corresponding to long wavelength density disturbances or long-scale migration. An algorithm for the efficient simulation of quantum nanosystems is suggested.

## I    Introduction

Quantum nanosystems are assemblies of thousands or more of fermions or bosons. Examples include quantum dots[1], superconducting nanoparticles[2], and graphene platelets and ribbons[3]. Droplets of 3He and 4He also display quantum behavior, and for similar reasons, Li as a confined quantum gas[4] or in a polymeric matrix[5]. Due to the size of quantum nanosystems, their behavior lies somewhere between that of small molecules and macroscopic systems. The objective of this study is to develop a theoretical framework that addresses the multiscale character of these systems.

Strongly interacting quantum many-particle systems cannot readily be analyzed via classic perturbation or path integral methods. The difficulty stems from the coupling of processes across multiple space-time scales and the strength of the inter-particle forces that underlie their dynamics. Multiscale analysis of the Schrödinger equation has been used to derive coarse-grained wave equations (CGWE) for boson[6] and fermion[7-8] systems. For the latter, the lowest order solution to the Schrödinger equation was taken to be in the form of an antisymmetric factor (for the ground state) times a long space-time scale factor for low-lying excitations. The latter factor was taken to be symmetric with respect to the exchange of the labels of identical fermions and therefore describes boson-like excitations.

Multiscale Analysis was also shown to imply that quantum nanosystems can satisfy a CGWE of a form reminiscent of the Dirac equation for the relativistic electron[8]. These results follow from the Schrödinger equation and the existence of distinct spatial scales inherent to those systems. The ratio $\varepsilon$ of the average nearest-neighbor distance to the size of the nanosystem or other nanometer-scale characteristic length enabled a perturbation scheme that lead to the CGWE.

Earlier multiscale treatments of quantum nanosystems left some ambiguity in the spatial scale at which coarse-graining is to be applied. One objective of the present treatment is to more directly relate the choice of scaling of space to that arising in the various contributions to the $N-$particle potential; this scale may be shorter or similar to the size of the nanoassembly depending on the physics of the constituent particles and, notably, the strength of their interactions.

Renormalization group methods have been used to understand extensive fermion systems and the emergence of effective masses and other modified quantities[9]. This technique addresses the fact that in a very large system there are many scales. The formalism is based on scaling transformations which, when repeated many times, leave the underlying equations invariant to further coarse-graining. However, for a nanosystem, a multiscale approach often yields an excellent approximation, e.g., for the classical Liouville[10-19] and the Poisson-boltzmann[20] equations.

The existence of a gap in timescales between that of individual atomistic fluctuations and the larger-scale coordinated motion of multiple atoms has been demonstrated for classical systems[17-19]. Furthermore, the order parameter momentum autocorrelation function was shown to have a short decay time; this implies that the order parameters decouple from other intermediate timescale processes[18-19]. With this and a multiscale perturbation method, a Langevin equation of stochastic order parameter dynamics was derived. These results enable the simulation of macromolecules and their complexes over long times (e.g., microseconds) which is impractical for these supramillion atom systems via traditional molecular dynamics simulations[18-19]. The implication of these classical mechanical results is that for some nanosystems one may use an approach based on a few discrete length and time scales. Here, we explore this notion for quantum nanosystems, avoiding the need for the complexity of the renormalization group approach.

A two-scale theory starts with the introduction of the position $\vec{r}_\ell$ of particle $\ell$ (in units where the average nearest-neighbor distance is one) and a scaled position $\vec{R}_\ell = \varepsilon \vec{r}_\ell$ (in terms of which traversal across a nano-assembly or across other long characteristic length involves a distance of one unit). Here, $\varepsilon$ is the ratio of the average nearest-neighbor distance to the long characteristic length (e.g., 100 nanometers). The basis of the multiscale analysis that we develop is that the wavefunction depends on $\vec{r}_\ell$ both directly and, through $\vec{R}_\ell$, indirectly. Although we pursue this two-scale approach in the present study, our methodology applies to cases with a range of intermediate scales, e.g., involving the dependence of the wavefunction on a set of scaled positions $\vec{R}_\ell^{(\gamma)} = \varepsilon^\gamma \vec{r}_\ell$, where $\gamma$ $(\gamma > 0)$ is a space-scaling exponent. However, since the two-scale approach was successful in establishing and implementing an efficient and accurate algorithm for simulating supramillion atom classical nanosystems (e.g., nonenveloped viruses and RNA)[18-19], in the present study we consider it for quantum nanosystems.

The two-scale approach presented here for low-lying excitations in fermion systems has analogies to methods used in porous media. In the latter case, equations of fluid flow on the pore scale are shown, after coarse-graining, to lead to Darcy flow, i.e., overall flow driven by a coarse-grained pressure gradient[21]. Such results follow from an expansion in the ratio of the pore size to the characteristic length of the overall pressure gradient or of large scale non-uniformities in the system. In more close analogy to the present problem, a two-scale approach shows that Darcy flow on one scale yields, upon coarse-graining, to Darcy flow on a larger scale but with modified permeability[22]. These findings have analogy in the present problem wherein coarse-graining of the wave equation for the $N-$fermion system leads to a CGWE with modified masses, interactions, and other characteristics.

The present study also explores the hypothesis that since the CGWE describes the interaction of each particle along a trajectory spanning many average nearest-neighbor distances, the CGWE yields solutions that, to good approximation, are factorizable, i.e., have mean-field character. This is distinct from a traditional mean-field theory. The latter follows from assuming a Slater determinant of single-particle functions and deriving equations for them via the variational principle using the original Hamiltonian. The present approach, however, accounts for particles with modified masses interacting with modified forces where equations are derived via the variational principle using the coarse-grained Hamiltonian.

The present theory is cast in a fully dynamical framework. Thus, stationary states corresponding to low-lying excitations and responses to time-dependent external fields can be studied. For example, the dynamical response to an applied electrical or magnetic field can be simulated in a study of graphene or nanowires of superconducting materials.

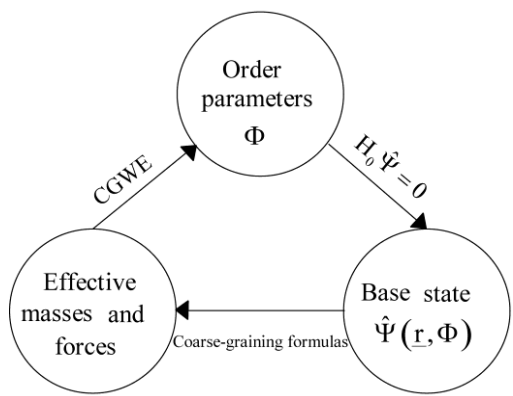

**Fig. 1**
Multiscale analysis of the wave equation captures the interplay of short and long space time dynamics.

The multiscale approach starts with the identification of a set of order parameters $\Phi$ characterizing nanoscale system features. These parameters (e.g., mass density disturbances) characterize the coherent behavior of many particles simultaneously or the long-range migration of single particles. As such, they evolve slowly in time and underlie the timescale separation that enables a multiscale analysis. In particular, the wave equation and the order parameters imply the existence of a timescale ratio $\eta$, e.g., the typical time of single particle collisions to that of collective oscillations or long length-scale migration. A key step in the multiscale analysis is to relate $\eta$ to the length-scale ratio $\varepsilon$ to provide a unified perturbation parameter, taken to be $\varepsilon$ here. The theory proceeds via a perturbation analysis in $\varepsilon$ to capture the interplay of processes across scales in space and time (Fig.1). To lowest order in $\varepsilon$, the solution to the wave equation takes the form of an antisymmetrized sum of terms like $\hat{\Psi}(\underline{r},\Phi)W(\Phi,\tilde{t})$ where $\hat{\Psi}$ is an eigenstate of the lowest order problem and $W(\Phi,\tilde{t})$ is a long-scale envelope-like factor that evolves with time $\tilde{t} = \varepsilon^{\tau}t$ for scaling exponent $\tau$. The case $\tau = 1$ was shown earlier to yield CGWE behaviors similar to those of the Dirac equation for the relativistic electron (believed to describe graphene [3]). In the present study we consider a distinct "universality class", i.e., for $\tau = 2$.

In this paper, a multiscale framework for strongly-interacting fermion nanosystems is set forth (Sect II) and a CGWE is derived (Sect III). The multiscale mean-field approximation for bosonic excitations is investigated in Sect IV, various families of excitations are discussed in Sect V, a crude estimate of the effective mass factors in the CGWE is given and discussed in Sect VI, and a multiscale computational algorithm is suggested in Sect VII. Conclusions are drawn in Sect VIII.

## II  The Multiscale Framework

Consider a system of $N$ identical fermions such as the electrons in a quantum dot or superconducting nanoparticle. Such systems display structure on at least two length-scales, e.g., the average nearest-neighbor distance and the overall size of the assembly. The objective of the following development is to transform the Schrödinger equation into a form that reveals the multiple scale character of the wavefunction. First, let the system be described by the $N$-particle configuration $\underline{r} = \{\vec{r}_1, \cdots \vec{r}_N\}$. In the original form, the wavefunction $\Upsilon(\underline{r},t)$ satisfies

$$i\hbar \, \partial \Upsilon / \partial t = H\Upsilon \tag{II.1}$$

where $H$ is the Hamiltonian

$$H = -\frac{\hbar^2}{2m}\nabla^2 + V \tag{II.2}$$

where $m$ is the mass of one of the $N$ identical fermions, $\nabla^2$ is the $3N$–dimensional Laplacian for the $\underline{r}$ space, and $V$ is the potential energy operator.

The wavefunction for a nanosystem has multiple possible dependencies on $\underline{r}$ and $t$. To express the multiscale complexity, we introduce an "unfolded wavefunction" $\Psi$ via

$$\Upsilon(\underline{r},t) = \Psi\left(\underline{r}, \underline{R}(\underline{r}), t_0(t), \underline{t}(t); \varepsilon\right) \tag{II.3}$$

where $\underline{R}(\underline{r})$ is a scaled configuration, i.e., $\underline{R} = \varepsilon^\sigma \underline{r}$ for scaling exponent $\sigma$; the set of times $\{t_0, \underline{t}\} = \{t_0, t_1, t_2, \ldots\}$ is related to $t$ via $t_n = \varepsilon^{\tau_n} t$ for a set of scaling exponents $\tau_0, \tau_1, \ldots$; and the smallness parameter $\varepsilon$ is a factor in the $N$–particle potential $V$ (it is the ratio of constants in the potential that weighs the contributions to the $N$–fermion potential from various forces). For concreteness, we consider the case $V = V_0 + \varepsilon^2 V_2$ to suggest that $V$ has two contributions $V_0$ and $\varepsilon^2 V_2$, essentially as a way to define $\varepsilon$. Then the analysis is found to imply that one may self-consistently choose the $\sigma$ and $\tau$ exponents to be integers. If this scaling had been inconsistent, it will lead to contradictions in the development which we take as an indicator to trigger a modification to the scaling ansatz. Finally, we have assumed that there are only two essential length scales, configurations $\underline{r}$ and $\underline{R} = \varepsilon^\sigma \underline{r}$. However, in principle, there can be multiple essential configurational variables $\underline{R}^{(n)} = \varepsilon^{\sigma_n} \underline{r}$ for a set of scaling exponents $\sigma_1, \sigma_2, \ldots$. If there are multiple particle types (e.g., electrons and ion cores in a metal), then the natural spatial scaling for various particle types may vary. For more complex systems the scaling approach must be modified. Here we focus on systems of $N$ identical fermions and phenomena involving two spatial scales such that $\sigma$ can be chosen to be an integer. Similarly, $\tau_n$ are chosen such that $\tau_n = n$. However, generalizations for more complex cases follow via the same logic; if inconsistencies emerge in the development, then the scaling exponents are modified accordingly.

The splitting of the potential $V$ into terms of various orders in $\varepsilon$ (i.e., $V = V_0 + \varepsilon V_1 + \varepsilon^2 V_2$) affects the structure of scaling limit laws. Starting with a physical rational for the splitting and a specification of the type of phenomenon to be studied, our objective is to provide a rational for the scaling of space and time and thus the resulting CGWE.

Consider a two-body potential $v(r)$ that depends on interparticle distance $r$. Let $G(r)$ be a Gaussian-like function (i.e., $G(0) = 1$ and $G \to 0$ as $r \to \infty$). Then $v$ can be split via

$$v(r) = v(r)G(r) + v(r)[1 - G(r)]. \tag{II.4}$$

The first term has short-scale character, i.e., approaches zero as $r \to \infty$ faster than $v(r)$. However, the second term is well-behaved as $r \to 0$ assuming by construction that $1 - G$ decays to zero as $r \to 0$ faster than $v(r) \to \infty$. By choice of the range over which $G$ decays, the second term can be considered a perturbation, i.e., contributes to $\varepsilon V_1$ rather than to $V_0$. For Coulomb systems such as electrons in a metal or semi-conductor, there is an additional reason why the $v(r)[1-G(r)]$ term can be considered a perturbation. As the distance between two electrons exceeds a few lattice spacing, the total potential (electron-electron plus electron-ion core) acts to screen the interaction; thus, each electron can be considered to be near an oppositely charged ion core so that $v(r)[1-G(r)]$ summed over the electrons and its neighboring ion core is a screened interaction.

In the present splitting ansatz, $V_0$ is independent of the set $\underline{R}$ of long-scale variables. Thus, the eigenstates of $H_0$, the $O(\varepsilon^0)$ contribution to the Hamiltonian, are independent of $\underline{R}$, the collection of long-scale particle position variables. If the potential has $\varepsilon$ terms (i.e., $V_1 \neq 0$), then the $O(\varepsilon)$ contribution to the CGWE will be shown to have solutions proportional to $\exp(i\langle 0|V_1|0\rangle t_1/\hbar)$ (where the bra-ket notation $|0\rangle$ corresponds to $\hat{\Psi}$, a stationary state of $H_0$) and hence oscillates with a different frequency at each point in $\underline{R}-$space (see Sect III). Such solutions eventually develop large gradients which is inconsistent with the fact that a coarse-grained wavefunction depends only on the set of long-scale variables $\underline{R}$. In the following section, we show that these considerations imply the choice of the scaling of space and time.

## III  Emergence of the Coarse-Grained Wave Equation (CGWE)

With the potential splitting and the scaling introduced in Sect II, we now use a two-scale approach to derive a CGWE for a system of $N$ identical fermions. To capture the long-scale dynamics, we introduce a set of additional variables $\underline{R} = \{\vec{R}_1, \ldots \vec{R}_N\}$ where $\vec{R}_\ell \equiv \varepsilon \vec{r}_\ell$ and $\vec{r}_\ell$ is the position of the $\ell^{\text{th}}$ particle. The $\underline{r}$-dependence tracks variations on the scale of the average nearest-neighbor spacing and hence individual particle-particle interactions, and thereby the exclusion principle which dominates short-scale dynamics. On the other hand, the $\underline{R}$-dependence reflects variations in $\Psi$ due to long-range correlation or migration across the $N$-particle assembly. Note that $\underline{R}$ is not a distinct set of dynamical variables; rather, $\Psi$ depends on $\underline{r}$ both directly and, through $\underline{R}$, indirectly. Similarly, we hypothesize that $\Psi$ also depends on time $t$ both directly and, through a set of scaled times, $\{t_0, \underline{t}\} = \{t_0, t_1, t_2, \ldots\}$ for $t_n = \varepsilon^n t$, indirectly where $t_0 = t$ tracks the shortest timescale changes and $t_n$ ($n > 0$) tracks the longer scale ones. In what follows, we show that these scaling hypotheses yield a self-consistent picture for small $\varepsilon$.

For a quasi-spherical nanoassembly constituted of $N$ fermions, $\varepsilon$ is $\mathrm{O}(N^{-1/3})$ so that one expects a multiscale perturbation expansion in $\varepsilon$ should be rapidly converging for assemblies of greater than 1000 particles. More generally, $\varepsilon$ is $\mathrm{O}(N^{-1/d})$ where $d$ is the dimensionality of the structure.

The multiscale hypothesis (II.3) and the chain rule imply the unfolded wave equation

$$i\hbar \sum_{n=0}^{\infty} \varepsilon^n \frac{\partial \Psi}{\partial t_n} = \left( H_0 + \varepsilon H_1 + \varepsilon^2 H_2 \right) \Psi \tag{III.1}$$

$$H_0 = \frac{-\hbar^2}{2m} \nabla_0^2 + V_0(\underline{r}) \tag{III.2}$$

$$H_1 = \frac{-\hbar^2}{m}\underline{\nabla}_0 \cdot \underline{\nabla}_1 \tag{III.3}$$

$$H_2 = \frac{-\hbar^2}{2m}\nabla_1^2 + V_2(\underline{r},\underline{R}) \tag{III.4}$$

where $V$ is written as $V_0 + \varepsilon^2 V_2$ (via the potential splitting discussed in Sect. II), while $\underline{\nabla}_0$ and $\underline{\nabla}_1$ are the $\underline{r}$ and $\underline{R}$ gradients, respectively. In other words, $\nabla_0^2 = -\frac{1}{\hbar^2}\sum_{\ell=1}^N \vec{p}_\ell \cdot \vec{p}_\ell$, $\nabla_1^2 = -\frac{1}{\hbar^2}\sum_{\ell=1}^N \vec{P}_\ell \cdot \vec{P}_\ell$, and $\underline{\nabla}_0 \cdot \underline{\nabla}_1 = -\frac{1}{\hbar^2}\sum_{\ell=1}^N \vec{p}_\ell \cdot \vec{P}_\ell$ where $\vec{p}_\ell = -i\hbar\,\partial/\partial\vec{r}_\ell$ at constant $\underline{R}$, $\vec{r}_{\ell'\neq\ell}$ and $\vec{P}_\ell = -i\hbar\,\partial/\partial\vec{R}_\ell$ at constant $\underline{r}, \vec{R}_{\ell'\neq\ell}$. In what follows, we solve (III.1) via a perturbation expansion in $\varepsilon$ (i.e., $\Psi = \sum_{n=0}^{\infty}\varepsilon^n \Psi_n$) and a solution to the wave equation is constructed at each order.

To $O(\varepsilon^0)$, the unfolded wave equation (III.1) takes the form

$$i\hbar\frac{\partial \Psi_0}{\partial t_0} = H_0 \Psi_0. \tag{III.5}$$

In the following, we seek excitations that are low lying disturbances from the ground state of $H_0$, denoted $\hat{\Psi}$ (assumed to be unique); however, more general solutions can also be considered. Adopting the convention that the ground state energy is zero (i.e., $H_0\hat{\Psi} = 0$), (III.5) admits a solution in the form

$$\Psi_0 = \hat{\Psi}(\underline{r})W(\underline{R},\underline{t}) \tag{III.6}$$

where the factor $W$ is found to be related to the lowest order coarse-grained wavefunction and is determined at higher order of the $\varepsilon$ development.

To $O(\varepsilon)$, the unfolded wave equation implies

$$i\hbar\left(\frac{\partial \Psi_1}{\partial t_0} + \frac{\partial \Psi_0}{\partial t_1}\right) = H_0\Psi_1 + H_1\Psi_0. \tag{III.7}$$

This admits the solution

$$\Psi_1 = S(t_0)\Psi_1^0 - \int_0^{t_0} dt_0' S(t_0 - t_0')\left\{\frac{\partial \Psi_0}{\partial t_1} + \frac{i}{\hbar}H_1\Psi_0\right\} \tag{III.8}$$

where $\Psi_1^0(\underline{r},\underline{R},\underline{t})$ is the initial value of $\Psi_1$ (i.e., at $t_0 = 0$) and $S(t_0)$ denotes the evolution operator $\exp(-i(H_0 - i0^+)t_0/\hbar)$. The positive infinitesimal $0^+$ is introduced to insure the evolution operator vanishes when $t_0 \to \infty$; this has implications for the concept of effective mass (see below).

Redefining $t_0'$ and inserting (III.3) and (III.6) in (III.8) yields

$$\Psi_1 = S(t_0)\Psi_1^0 - t_0|0\rangle\frac{\partial W}{\partial t_1} - \frac{i}{\hbar}\int_{-t_0}^{0} dt_0' S(-t_0')H_1(W|0\rangle) \tag{III.9}$$

where the bra-ket notation $|0\rangle$ is used to represent $\hat{\Psi}$.

To further the analysis, we use a theorem analogous to the Gibbs hypothesis stating that the long-time average and expectation value are equal (Appendix A):

$$\lim_{t_0 \to \infty}\frac{1}{t_0}\int_{-t_0}^{0} dt_0' S(-t_0')\Omega|0\rangle = |0\rangle\langle 0|\Omega|0\rangle \tag{III.10}$$

$$\langle 0|\Omega|0\rangle \equiv \int d^{3N}r\,\hat{\Psi}^*\Omega\hat{\Psi} \tag{III.11}$$

for any time-independent operator $\Omega$.

Examination of (III.9) shows that for $\Psi_1$ to be well-behaved the $t_0$-divergent terms must be counterbalanced as $t_0 \to \infty$, or, if there are no such counterbalancing terms, then $W$ must be independent of $t_1$. Taking the inner product with $\langle 0|$ on both sides, multiplying by $1/t_0$ and letting $t_0 \to \infty$, one obtains

$$i\hbar \frac{\partial W}{\partial t_1} = -\frac{i\hbar}{m} \sum_{\ell=1}^{N} \langle 0|\vec{p}_\ell|0\rangle \cdot \frac{\partial W}{\partial \vec{R}_\ell}. \tag{III.12}$$

Since $H_0|0\rangle = 0$ and thus so is $\langle 0|H_0$, the uniqueness of $\hat{\Psi}$ for the non-degenerate case implies $\hat{\Psi}$ can be taken as real without loss of generality. With this, $\langle 0|\vec{p}_\ell|0\rangle = (i/2\hbar)\int d^{3N}r\, \partial \hat{\Psi}^2/\partial \vec{r}_\ell$. Since $\hat{\Psi}$ vanishes at infinity for a bounded system, one finds that $\langle 0|\vec{p}_\ell|0\rangle = 0$. With this, we conclude that $W$ is independent of $t_1$ (i.e., $\partial W/\partial t_1 = 0$). Therefore, (III.9) becomes

$$\Psi_1 = S(t_0)\Psi_1^0 - \frac{i}{\hbar}\int_{-t_0}^{0} dt_0' S(-t_0') H_1(|0\rangle W). \tag{III.13}$$

To $O(\varepsilon^2)$, the wave equation (III.1) implies

$$i\hbar\left(\frac{\partial \Psi_2}{\partial t_0} + \frac{\partial \Psi_1}{\partial t_1} + \frac{\partial \Psi_0}{\partial t_2}\right) = H_0 \Psi_2 + H_1 \Psi_1 + H_2 \Psi_0. \tag{III.14}$$

This admits the solution

$$\Psi_2 = S(t_0)\Psi_2^0 - \int_{-t_0}^{0} dt_0' S(-t_0')\left\{\frac{\partial \Psi_0}{\partial t_2} + \frac{\partial \Psi_1}{\partial t_1} + \frac{i}{\hbar}H_1\Psi_1 + \frac{i}{\hbar}H_2\Psi_0\right\} \tag{III.15}$$

where $\Psi_2^0(\underline{r},\underline{R},\underline{t})$ is the value of $\Psi_2$ at $t_0 = 0$. In what follows, we consider the class of initial data to be in the form $\hat{\Psi}W$ (for further discussion, see earlier work on Newtonian systems[13-14]), i.e., $\Psi_n^0 = 0$ for $n \geq 1$. Inserting (III.3), (III.4), (III.6), and (III.13) in (III.15) yields

$$\Psi_2 = -t_0|0\rangle\frac{\partial W}{\partial t_2} + t_0\frac{i\hbar}{2m}|0\rangle\nabla_1^2 W - \frac{i}{\hbar}\int_{-t_0}^{0} dt_0' S(-t_0')V_2|0\rangle W \tag{III.16}$$

$$+\frac{i}{\hbar}\int_{-t_0}^{0}\int_{-t_0}^{0} dt_0' dt_0'' S(-t_0')S(-t_0'')H_1\left(|0\rangle\frac{\partial W}{\partial t_1}\right)$$

$$-\frac{\hbar^2}{m^2}\int_{-t_0}^{0}\int_{-t_0}^{0} dt_0' dt_0'' S(-t_0')\underline{\nabla}_0 \cdot \underline{\nabla}_1 \{S(-t_0'')\underline{\nabla}_0 \cdot \underline{\nabla}_1(|0\rangle W)\}.$$

The condition guaranteeing $\Psi_2$ is well-behaved as $t_0 \to \infty$ yields

$$i\hbar \frac{\partial W}{\partial t_2} = H^{CG} W \quad \text{(III.17)}$$

$$H^{CG} = V^{CG} + \sum_{\ell,\ell'=1}^{N} \sum_{\alpha,\alpha'=1}^{3} \mu_{\ell\alpha\ell'\alpha'} \frac{\partial^2}{\partial R_{\ell\alpha} \partial R_{\ell'\alpha'}} \quad \text{(III.18)}$$

$$V^{CG} = \langle 0 | V_2 | 0 \rangle \quad \text{(III.19)}$$

$$\mu_{\ell\alpha\ell'\alpha'} = -\frac{\hbar^2}{2m} \delta_{\ell\ell'} \delta_{\alpha\alpha'} + \tilde{\chi}_{\alpha\alpha'} \quad \text{(III.20)}$$

$$\tilde{\chi}_{\alpha\alpha'} = \frac{i\hbar}{m^2} \int_{-\infty}^{0} dt_0 \chi_{\ell\alpha\ell'\alpha'}(t_0) \quad \text{(III.21)}$$

$$\chi_{\ell\alpha\ell'\alpha'}(t_0) = \langle 0 | p_{\ell\alpha} S(-t_0) p_{\ell'\alpha'} | 0 \rangle \quad \text{(III.22)}$$

(see Appendix B for further discussion). The $\mu$ term corresponds to an effective inverse mass that is tensorial in character and is a two-body term (i.e., depends on $\ell$ and $\ell'$).

To this point we have not explored the physical interpretation of $W$. Define a coarse-grained wavefunction $\Psi^{CG}$ via

$$\Psi^{CG} = \frac{\int d^{3N} r \, \Delta(\underline{R} - \varepsilon \underline{r}) \hat{\Psi}^* \Psi}{\int d^{3N} r \, \Delta(\underline{R} - \varepsilon \underline{r}) |\hat{\Psi}|^2} \quad \text{(III.23)}$$

where $\Delta(\underline{R} - \varepsilon \underline{r})$ is a function that is narrowly distributed around $\underline{0}$ and is zero otherwise. Placing the series for $\Psi$ (i.e., $\Psi = \Psi_0 + \varepsilon \Psi_1 + \cdots$) and the expression for $\Psi_0$ from (III.6) in (III.23), one finds $\Psi^{CG} = W$ when $\varepsilon \to 0$. With this, (III.17) can be considered a CGWE describing the long space-time dynamics of low-lying excitations of an $N$−fermion system.

The lowest order solution $\Psi_0$ has the form $\hat{\Psi}(\underline{r}) W(\underline{R}, \underline{t})$. At first sight, this seems to imply that particle $\ell$ $(\ell = 1, 2, \ldots N)$ can be at two places at the same time, i.e., $\vec{r}_\ell$ and $\vec{R}_\ell$. Rather, the correct interpretation is that the lowest order solution is $\hat{\Psi}(\underline{r}) W(\varepsilon \underline{r}, \underline{t})$, i.e., that the $\underline{R}$-dependence in $W$ provides a slowly varying envelope over-and-above the highly fluctuating factor $\hat{\Psi}(\underline{r})$. For example, the probability density for $\varepsilon \to 0$ has the form $|\hat{\Psi}(\underline{r})|^2 |W(\underline{R},\underline{t})|^2$. While the $|\hat{\Psi}(\underline{r})|^2$ factor expresses the detailed short-scale structure of the system, $|W(\underline{R},\underline{t})|^2$ modulates this probability density with an overall envelope. With this, the expectation of an operator $\Omega$ is given by

$$\langle \Omega \rangle = \int d^{3N}r\, \hat{\Psi}^*(\underline{r}) W^*(\varepsilon \underline{r},\underline{t}) \Omega \left[ \hat{\Psi}(\underline{r}) W(\varepsilon \underline{r},\underline{t}) \right]. \tag{III.24}$$

For example, the expectation of the momentum of particle $\ell$, $\vec{\mathcal{P}}_\ell = -i\hbar \partial/\partial \vec{r}_\ell$, is given by

$$\langle \vec{\mathcal{P}}_\ell \rangle = -i\hbar \int d^{3N}r\, \hat{\Psi}^*(\underline{r}) W^*(\varepsilon \underline{r},\underline{t}) \left( \frac{\partial \hat{\Psi}}{\partial \vec{r}_\ell} W + \varepsilon \hat{\Psi} \frac{\partial W}{\partial \vec{R}_\ell} \bigg|_{\underline{R}=\varepsilon \underline{r}} \right). \tag{III.25}$$

In this case, the envelope factor contributes a $O(\varepsilon)$ term. However, the full contribution of this term must be evaluated using $\Psi_1$ as well.

In developing the multiscale framework, recall that the scaled particle position configuration $\underline{R} = \varepsilon \underline{r}$ does not mean that $\underline{R}$ is an additional set of variables (i.e., there are not $6N$ degrees of freedom). In deriving the implications of the theory one must return to the full description (i.e., $\Psi(\underline{r},\varepsilon \underline{r};t,\varepsilon t,\ldots;\varepsilon)$) and then arrive at the physical quantities. Thus, the expectation of an operator $\Omega$ is given by $\langle \Omega \rangle = \int d^{3N}r\, \Psi^*(\underline{r},\varepsilon \underline{r};t,\varepsilon t,\ldots;\varepsilon) \Omega \Psi(\underline{r},\varepsilon \underline{r};t,\varepsilon t,\ldots;\varepsilon)$. Also, one must re-establish the normalization condition $\int d^{3N}r\, |\Psi(\underline{r},\varepsilon \underline{r};t,\varepsilon t,\ldots;\varepsilon)|^2$, and similarly for other quantities. The multiscale framework provides a taylor series approximation in $\varepsilon$ with which these expressions may be evaluated.

# IV    Coarse-Grained Mean-field (CGMF) Approximation: Bosonic Excitations

The CGWE of Sect. III describes the longer space-time scale dynamics of a fermion nanosystem. Evolution is on the $\varepsilon^{-2}$ timescale and the $\varepsilon^{-1}$ length scale. Since $\Psi_0$ is antisymmetric, there are three possible cases.

Case I: $\hat{\Psi}$ is antisymmetric (fermionic) while $W$ is symmetric (bosonic).

Case II: $\hat{\Psi}$ is symmetric (bosonic) while $W$ is antisymmetric (fermionic).

Case III: An entangled case where antisymmetry is not attained as a simple product of $\hat{\Psi}$ and $W$.

A fermionic $W$ (case II) corresponds to a collection of $N$ interacting fermion-like quasi-particles with modified forces and effective masses. However, since $W$ depends on $\underline{R}$ and not $\underline{r}$, it cannot support the richness of short-scale variations needed to satisfy the exclusion principle. Thus, Case II is unphysical. Case III will be discussed in the next section.

As $W$ is a long-scale function (i.e., depends on $\underline{R}$ and $\underline{t}$, and not $\underline{r}$ and $t_0$), it must, to good approximation, have mean-field character since each particle interacts with many others in the large intervals of space-time described by $W$. Thus, for case I (bosonic $W$), $W$ can be approximated by a symmetrized product of single-particle functions. In the following, we consider a coarse-grained mean-field (CGMF) approximation to the CGWE.

*Bose Condensate*

The CGWE supports a Bose condensate wherein all quasi-particles are in the same state. For stationary states of this type, the ground state (denoted $W^{(0)}$) is given by

$$W^{(0)}(\underline{R}) = \prod_{\ell=1}^{N} A(\bar{R}_\ell) \tag{IV.1}$$

for single-particle function $A(\vec{R})$. The $W^{(0)}$ excitations correspond to all particles being similarly elevated in energy so that the energy of excitations of that type should be proportional to $N$.

In the CGMF approximation as formulated here, the single particle functions are determined via the variational principle. Let $E^{CG}$ be the coarse-grained excitation energy relative to the ground state. The stationary states of (III.17) satisfy $H^{CG}W = E^{CG}W$. With this, the stationary single particle functions are the extrema of the functional $\tilde{E}$ defined via

$$\tilde{E}[W] = \frac{\int d^{3N}R \, W^* H^{CG} W}{\int d^{3N}R \, |W|^2}. \tag{IV.2}$$

The single-particle function $A$ (assumed real for simplicity here) associated with $W^{(0)}$ is determined by

$$\frac{\delta \tilde{E}}{\delta A(\vec{R})} = 0. \tag{IV.3}$$

Taking the coarse-grained potential $V^{CG}$ to be a sum over two-body terms

$$V^{CG} = \frac{1}{2} \sum_{\ell=1}^{N} \sum_{\ell' \neq \ell=1}^{N} v(\vec{R}_\ell, \vec{R}_{\ell'}), \tag{IV.4}$$

(IV.3) implies

$$\left\{ Q(\vec{R}) + \sum_{\alpha,\alpha'=1}^{3} \tilde{\mu}_{\alpha\alpha'} \frac{\partial^2}{\partial R_\alpha \partial R_{\alpha'}} \right\} A(\vec{R}) = \tilde{E}' A(\vec{R}) \tag{IV.5}$$

$$\tilde{\mu}_{\alpha\alpha'} = \frac{-\hbar^2}{2m} \delta_{\alpha\alpha'} + \tilde{\chi}_{\alpha\alpha'} \tag{IV.6}$$

$$Q(\vec{R}) = (N-1) \int d^3R' A(\vec{R}')v(\vec{R},\vec{R}')A(\vec{R}') \tag{IV.7}$$

$$\tilde{E}' = \tilde{E} - (N-1) \int d^3R' A(\vec{R}') \sum_{\alpha,\alpha'=1}^{3} \tilde{\mu}_{\alpha\alpha'} \frac{\partial^2}{\partial R'_\alpha \partial R'_{\alpha'}} A(\vec{R}') \tag{IV.8}$$

$$- (N-1)(N-2) \int d^3R' d^3R'' \, A(\vec{R}')A(\vec{R}'')\hat{q}(\vec{R}',\vec{R}'')A(\vec{R}')A(\vec{R}'')$$

$$\hat{q}(\vec{R}',\vec{R}'') = \frac{1}{2}v(\vec{R}',\vec{R}'') + \sum_{\alpha,\alpha'=1}^{3} \tilde{\chi}_{\alpha\alpha'} \frac{\partial^2}{\partial R'_\alpha \partial R''_{\alpha'}}. \tag{IV.9}$$

This constitutes a nonlinear eigenvalue problem for determining $A(\vec{R})$ and $\tilde{E}$. This equation is distinct from that of traditional mean-field theory in that the kinetic energy term has two-body character and is tensorial in nature, the forces are coarse-grained, and the excitations are bosonic and not fermionic.

*Single and Multiple Quasi-Boson Excitations*

Let $\Xi_+$ be an operator that symmetrizes any function of $\underline{R}$. In the present case, the function to be symmetrized is a product of single particle functions. For example,

$$W^{(1)} = \Xi_+ B(\vec{R}_1) A(\vec{R}_2) \cdots A(\vec{R}_N), \tag{IV.10}$$

where $A$ and $B$ are single-particle functions. There are $N$ terms in the symmetrized sum constituting $W^{(1)}$. For large $N$, $A$ satisfies IV.5 which is nonlinear and is independent of the single-particle excitation factor $B$. The mean-field interaction of $B$ only involves $A$ and represents the $(N-1)$-particle background. To insure that $A$ and $B$ are distinct, the minimization of $\tilde{E}$ is carried out constrained by orthogonality, i.e., $\int d^3 R A(\vec{R}) B(\vec{R}) = 0$. However, the distinguished particle is surrounded by many background particles, and thus the $B$-factor satisfies an equation reflecting that the exceptional particle evolves in the fluctuating medium of the majority, the latter described by $A$. The single-particle excitation factor $B$ obeys an equation involving $A$ which provides the environment for the single-particle motion. Furthermore, the equation for $B$ is linear. Thus, one may solve the $A$ equation and then use the results to determine $B$ for large $N$. By analogy, for more complex excitations, one may construct functions involving various numbers of distinct single particle functions.

## V  Antisymmetry and Families of Excitations

In the previous sections we introduced a scaling relationship between the characteristic distance and the strength of terms in the $N-$particle potential. While the development is found to be consistent, the result is not necessarily unique. This does not imply that there is a missing constraint on the physics. Rather, there may be distinct types of behaviors that are related to a given class of initial states of the system.

While the wave function for an assembly of $N$ identical fermions must be antisymmetric, there are solutions to the Schrödinger equation which are of mixed symmetry and are still relevant as follows. Let $\xi_s$ be an operator that performs a permutation $s$ which represents a rearrangement of the particle configuration $\{\vec{r}_1, \vec{r}_2 ..., \vec{r}_N\}$ to $\{\vec{r}_{s_1}, \vec{r}_{s_2} ..., \vec{r}_{s_N}\}$ and $\Xi = \left(1/\sqrt{N!}\right) \sum_{s=1}^{N!} (-1)^s \xi_s$ is an antisymmetrizer. For example, a Slater determinant can be written as $\Xi$ times a product of $N$ single particle functions. With this, one may solve the Schrödinger equation in a space of arbitrary exchange symmetry and, via $\Xi$, reassemble the solutions into an antisymmetric function.

By construction, $H_0$, $H_1$, and $H_2$ are symmetric operators so that $\Xi$ commutes with them. We seek long space-time excitations of the system constructed as disturbances from the ground state of $H_0$. The family of excitations is constructed starting from a "base" solution $\hat{\Psi}$, choosing the energy convention such that $H_0 \hat{\Psi} = 0$. To be consistent with the antisymmetry requirement, and to be sufficiently general, we write

$$\Psi_0 = \Xi\left(\hat{\Psi} W\right). \tag{V.1}$$

Henceforth, we pursue each family of excitations separately. A family is categorized according to the exchange symmetry of the base solution $\hat{\Psi}$. Since $\hat{\Psi}$ may not be antisymmetric, it can have energy lower than the antisymmetric ground state of $H_0$. However, as the symmetry of $\hat{\Psi}$ departs from antisymmetry in increasing degree, $W$ will have increasingly antisymmetric character and thus have higher energy, to compensate for it.

*Collective Bosonic versus Single-Particle Fermionic Excitations*

The multiscale analysis of Sect. III reveals the existence of several families of excitations in $N-$fermion systems. In the simplest case, $\hat{\Psi}$ is antisymmetric and $W$ depends on $\underline{R}$ in a symmetric fashion; these excitations, as manifested in $W$, have thus a bosonic and long-scale character. However, there are other excitations which have more fermionic character as follows.

Inserting (V.1) and (III.6) in (III.8) yields

$$\Psi_1 = -\Xi \left\{ t_0 |0\rangle \frac{\partial W}{\partial t_1} + \frac{i}{\hbar} \int_{-t_0}^{0} dt'_0 S(-t'_0) H_1 (W|0\rangle) \right\} \qquad (V.2)$$

where $\Psi_1^0$ is taken to be zero as discussed in Sect III. Examining the long time behavior of (V.2) implies

$$\Xi |0\rangle \frac{\partial W}{\partial t_1} = 0 . \qquad (V.3)$$

With this, continuing the analysis to $O(\varepsilon^2)$ yields

$$\Xi |0\rangle \left\{ i\hbar \frac{\partial W}{\partial t_2} - H^{CG} W \right\} = 0 . \qquad (V.4)$$

This implies

$$\left\{ i\hbar \frac{\partial}{\partial t_2} - H^{CG} \right\} |0\rangle W = F , \qquad (V.5)$$

where $F$ is a symmetric function. Since $|0\rangle W = \Psi_0$ is antisymmetric, (V.5) implies that either the operator $\{i\hbar \partial/\partial t_2 - H^{CG}\}$ is antisymmetric or $F$ is a constant. However, $\{i\hbar \partial/\partial t_2 - H^{CG}\}$ is symmetric and therefore the only valid solution is that $F = C$ where $C$ is a constant. Multiplying both sides by $\langle 0 |$ yields

$$\left\{ i\hbar \frac{\partial}{\partial t_2} - H^{CG} \right\} W = C \langle 0 |. \tag{V.6}$$

However, the left hand side of (V.6) have no $\underline{r}$ dependency while $\langle 0 |$ on the right hand side does. This implies C can only be zero. With this, we obtain

$$i\hbar \frac{\partial W}{\partial t_2} = H^{CG} W. \tag{V.7}$$

Thus, (V.7) generates a family of excitations characterized by $W$ which obeys a CGWE, factors within which differ for different solutions $\hat{\Psi}$. Note that when we write $H_0 \hat{\Psi} = 0$ for a given family, we assume that the ground state energy for that family has been subtracted from $V_0$. This shows that even for complex entangled excitations $W$ still obeys the same CGWE.

# VI    Simple Estimate for Bosonic Excitations

Rough estimates for the factors in the CGWE were made to gain insights into the mathematical development. In particular, the effective mass tensor of III.20 was generated using a decoupled-particle approximation. The $V_0$ contribution to the many-particle potential is, by construction, short-range. As a simple approximation, we assume that electrons evolving according to $V_0$ can be taken to occupy single particle-like orbitals localized to a lattice of ion cores. In this picture, the delocalized nature of the electrons will arise solely due to the CGWE and $W$.

In this spirit, one could construct $\hat{\Psi}$ (denoted $|0\rangle$), and similarly for the excited states $|n\rangle$ $(n>0)$ of $H_0$. With the above or other approximation to the states of $H_0$ one can compute the effective mass tensor $\mu_{\ell\alpha\ell'\alpha'}$ and coarse-grained potential $V^{CG}$. When computing the effective mass tensor, one must evaluate the time integral of the single-particle momentum correlation

$$\tilde{\chi}_{\alpha\alpha'} = \frac{\hbar^2}{m^2} \sum_{n \neq 0} \frac{\langle 0|p_\alpha|n\rangle\langle n|p_{\alpha'}|0\rangle}{\zeta_n}, \tag{VI.1}$$

for $H_0$ excitation energy $(\zeta_n > 0)$. To obtain a crude estimate, consider a single-particle picture wherein the states $|n\rangle$ are limited to the three lowest coulomb excited states, i.e., the degenerate states $2p_z$, $2p_x$, $2p_y$ denoted $|1\rangle$, $|2\rangle$, $|3\rangle$ respectively (the $2s$ excited state does not contribute since for it $\langle 0|\vec{p}_\ell|n\rangle$ vanishes). Thus, $\zeta_1 = \zeta_2 = \zeta_3 = 3\hbar^2/8ma^2$ where $a$ is the Bohr radius. With this, we obtain

$$\tilde{\chi}_{\alpha\alpha'} = \frac{2^{12}}{3^9} \frac{\hbar^2}{m} \delta_{\alpha\alpha'} = 0.2081 \frac{\hbar^2}{m} \delta_{\alpha\alpha'} \tag{VI.2}$$

$$\mu_{\ell\alpha\ell'\alpha'} = K_{\ell\ell'} \delta_{\alpha\alpha'} \tag{VI.3}$$

$$K_{\ell\ell'} = \left(0.2081 - 0.5\delta_{\ell\ell'}\right)\hbar^2/m \tag{VI.4}$$

With this, the coarse-grained Hamiltonian takes the form

$$H^{CG} = V^{CG} + \sum_{\ell,\ell'=1}^{N} K_{\ell\ell'} \left\{ \frac{\partial^2}{\partial R_{\ell 1} \partial R_{\ell' 1}} + \frac{\partial^2}{\partial R_{\ell 2} \partial R_{\ell' 2}} + \frac{\partial^2}{\partial R_{\ell 3} \partial R_{\ell' 3}} \right\} \tag{VI.5}$$

To interpret these results, we transform the particle coordinates such that the effective inverse mass tensor becomes diagonal. Thus, we introduce a set of modified positions

$$\tilde{R}_{\ell\alpha} = \sum_{\ell'=1}^{N} Q_{\ell\ell'} R_{\ell'\alpha} \tag{VI.6}$$

such that the matrix $Q$ satisfies the condition

$$QKQ^T = D \tag{VI.7}$$

where $D$ is a diagonal matrix. With this, (VI.5) takes the form

$$H^{CG} = V^{CG} + \sum_{\ell=1}^{N} d_\ell \nabla^2_{\tilde{R}_\ell} \tag{VI.8}$$

where $d_\ell$ is the $\ell^{th}$ diagonal entry of matrix $D$. Solving (VI.7) yields

$$d_\ell = \begin{cases} -0.2919 \dfrac{\hbar^2}{m} & \text{for } \ell = 1 \\ \dfrac{0.08521 - 0.06074(\ell-2) - 0.04331(\ell-1)}{-0.2919 + 0.2081(\ell-2)} \dfrac{\hbar^2}{m} & \text{otherwise} \end{cases} \tag{VI.9}$$

Notice that as $\ell$ increases, $d_\ell$ approaches $-\hbar^2/2m$, the bare mass value. Thus, associated with these excitations is an effective mass $m_{eff}$ such that $m \leq m_{eff} \leq 1.713m$. However, this is not simply interpreted as corresponding to a set of bosonic excitations with effective $m_{eff}$ since $W$ is a symmetric function of $R$ and not $\tilde{R}$.

The implications of the effective masses for excitations are more easily understood within the mean-field approximation of sect IV. In this case, the inverse mass tensor takes the value $\tilde{\mu}_{\alpha\alpha'} = -0.2919\hbar^2\delta_{\alpha\alpha'}/m$. For the ground state of the CGWE, $m_{eff}$ in the equation for the single particle function (IV.1) is equal to the largest value in the effective mass interval, i.e., $1.713m$. In general, it is expected that this value would be sensitive to lattice structure and the type of atoms constituting the metal or semiconductor. Thus, while this result is simply a crude estimate based on the hydrogenic wavefunctions, it illustrates that the coarse-graining induces second order terms in the CGWE which shifts the effective mass higher than the bare one.

# VII    Implementation as a Multiscale Computational Platform

The conceptual flow of Fig. 1 suggests an algorithm for the multiscale simulation of quantum nanosystems. Its implementation presents two challenges: (1) constructing $\hat{\Psi}$ and (2) computing the inverse mass tensor $\mu_{\ell\alpha\ell'\alpha'}$ and coarse-grained potential $V^{CG}$ appearing in the coarse-grained Hamiltonian $H^{CG}$. These quantities, and in particular $\hat{\Psi}$ needed to compute them, can be constructed via path integral[23-24], DFT, variational techniques. Given that $V_0$, by construction, expresses short-range interactions only, construction of $\hat{\Psi}$ is somewhat easier than for the full problem which, for a coulomb system, has long-range correlation.

For a variational approach to the entangled case (Sect V), we start with a trial function that has $N^*$ quasi-particles distinguishable via their long-scale behavior. The corresponding states $\hat{\Psi}$ are the analogue of a Slater determinant expression. Consider a trial function involving short-scale functions $\varphi_k(\vec{r})$ $(k=1,2,\ldots N)$ and long-scale ones $w_q(\vec{R})$ $(q=1,2,\ldots N^*+1)$. An antisymmetric function $\Psi_0$ can be written as

$$\Psi_0 = \Xi\big(\varphi_1(1)\cdots\varphi_N(N)\big)\big(w_1(1)\cdots w_{N^*}(N^*)w_{N^*+1}(N^*+1)\cdots w_{N^*+1}(N)\big). \tag{VII.1}$$

Here, $\varphi_k(\ell) \equiv \varphi_k(\vec{r}_\ell)$ and $w_q(\ell) \equiv w_q(\vec{R}_\ell)$.

The generalization of the above mean-field Slater determinants and their optimization via a variational approach allows for a spectrum of excitations. For example, $N^*=0$ makes $\hat{\Psi}$ antisymmetric and the analysis of Sect. III follows. The $N^*=1$ family corresponds to single quasi-particle excitations, and similarly for $N^*>1$. Finally a variational procedure can be used to optimize the single particle functions $\varphi_k(\vec{r})$. Also, the variational form (VII.1) can be augmented by multiplying it with a forfactor that is a symmetric function of $\underline{r}$ and which has variational parameters that can be used to account for some correlation in the short scale behavior $\left(\text{i.e.}, \hat{\Psi}\right)$.

## VIII  Conclusions

Long space-time behaviors of fermion nanosystems (e.g., quantum dots, superconducting nanoparticles, and graphene nanostructures) follow from the interaction of each fermion with many others across multiple scales in space and time. Thus, it is expected that the longer space-time dynamics of fermion nanosystems is inherently of mean-field character. When combined with a multiscale methodology for constructing a CGWE, this implies a coarse-grained mean-field picture of the low-lying excitations of fermion nanosystems. The resulting CGMF approximation suggests an algorithm for the efficient simulation of these low-lying excitations.

The collective, bosonic excitations supported by a many-fermion system are of several types. The simplest excitations, according to the CGMF picture, involve all bosonic excitations in the ground state (i.e., the condensate). If all these bosonic excitations are in the same state, then the excitation energy increases with $N$. Other excitations wherein the individual bosonic degrees of freedom are in different energy levels can have excitation energy relative to the condensate that is not proportional to $N$ (e.g., IV.10).

Quasi-particle excitations with more fermionic character were also investigated. These correspond to solutions of the multiscale wave equation which, as $\varepsilon \to 0$, are antisymmetrized sums of products of a short-scale core solution $\hat{\Psi}$ and a long-scale function $W$. Such particle-like excitations (e.g., quasi-particles in a semiconductor), emerge from an analysis wherein the long-scale variations are entangled with the short-scale ones to capture both the multiscale nature of the wavefuntion and respect the exclusion principle.

To realize the practical implications of the present theory, several advances must be made:

1) Construction of the factors in the CGWE requires an expression, or a very efficient computational algorithm, for constructing $\hat{\Psi}$.

2) Advancing the system in time via the CGWE could be facilitated via a type of path integral approach; this follows from the notion that construction of $\underline{R}$-trajectories could proceed in a discretized fashion involving larger time steps (e.g., due to the $t_2$ rather than the $t_0$ dependence of the dynamics). Thus, as $\underline{R}$ progresses along a trajectory, one needs only to compute the CGWE factors in the immediate vicinity of the instantaneous $\underline{R}$ values. Hence, the entire $\underline{R}$ dependence of these factors is not required. The flowchart for such a stepwise computation is suggested in Fig.1.

3) A variation approach can be used to construct $\hat{\Psi}$. This can be accomplished by (1) making an ansatz on the form of $\hat{\Psi}$ with a set of variational parameters $\lambda$, (2) using Monte Carlo integration to compute the value of the variational energy for given $\lambda$, and (3) using an efficient minimization algorithm to find the best value of $\lambda$. In the time-stepping used to construct the trajectories for the coarse-grained dynamics, the Monte Carlo integrations over $\underline{r}$ are limited to a $3N$ dimensional cube centered around $\varepsilon^{-1}\underline{R}$ and extend a limited distance over each of the $3N$ directions for an interval of length $b/2$ where $b^{3N}$ is the volume in $\underline{r}$-space over which $\Delta(\underline{R} - \varepsilon \underline{r})$ is nonzero.

The spectrum of a system like nanoscale rings, disks, and spheres reflects its geometry and the universality class of its long-scale behavior. Consider the CGMF approximation. Eigenstates of the non-linear mean-field equations satisfy periodic boundary conditions for these systems. The single-particle wavefunction $w$ (such as $A$ or $B$ of Sect IV) must satisfy the condition $w(X+L) = w(X)$ for a ring of circumference $L$. For a disk or sphere, $w$ satisfies analogous periodic conditions. With this, it is seen that the coarse-graining can express continuous symmetries not present in the original problem. For example, a metal or semi-conductor expresses the symmetry of the ion-core lattice while, in the CGWE, variations on this scale are absent.

In conclusion, the multiscale approach yields insight into the spectra of long space-time excitations of fermion nanosystems. It allows for the simulation of larger systems than can readily be achieved using standard methods such as implemented in Gaussian.

**Acknowledgements**

This project was supported in part by the National Science Foundation (Theory, Models and Computational Methods Program) and Indiana University's college of Arts and Sciences through the Center for Cell and Virus Theory.

**Appendix A: Long-time average/expectation value equivalence theorem**

The objective here is to show that the long-time average and expectation value are equal. First introduce the complete (infinite) set of eigenstates $|n\rangle$ of $H_0$ ($H_0|n\rangle = \zeta_n|n\rangle$ for energy $\zeta_n$ and $\zeta_0 = 0$). With this, for any operator $\Omega$ one has

$$\Omega|0\rangle = \sum_{n=0}^{\infty} |n\rangle\langle n|\Omega|0\rangle. \tag{A.1}$$

The long-time average of the time-evolved effect of $\Omega$ can be written as

$$\lim_{t_0 \to \infty} \frac{1}{t_0} \int_{-t_0}^{0} dt_0' S(-t_0')\Omega|0\rangle = \sum_{n=1}^{\infty} \langle n|\Omega|0\rangle \lim_{t_0 \to \infty} \frac{1}{t_0} \int_{-t_0}^{0} dt_0' S(-t_0')|n\rangle. \tag{A.2}$$

The long-time average of $S(-t_0')|n\rangle$ vanishes when $\zeta_n \neq 0$ and is equal to $|n\rangle$ otherwise (i.e., when $n=0$). Therefore, (A.2) becomes

$$\lim_{t_0 \to \infty} \frac{1}{t_0} \int_{-t_0}^{0} dt_0' S(-t_0')\Omega|0\rangle = |0\rangle\langle 0|\Omega|0\rangle, \tag{A.3}$$

the long-time average/expectation value equivalence theorem.

# Appendix B: The Correlation Function $\chi_{\ell\alpha\ell'\alpha'}(t_0)$ of (III.22)

The single-particle correlation function arising in (III.22) is of the form

$$\chi_{\ell\alpha\ell'\alpha'}(t_0) = \langle 0| p_{\ell\alpha} \exp\left(it_0\left(H_0 - i0^+\right)/\hbar\right) p_{\ell'\alpha'} |0\rangle. \tag{B.1}$$

Using the completeness of the eigenfunction $|n\rangle$ of $H_0$, this becomes

$$\chi_{\ell\alpha\ell'\alpha'}(t_0) = \sum_{n=0}^{\infty} \langle 0| p_{\ell\alpha} |n\rangle \langle n| p_{\ell'\alpha'} |0\rangle e^{iw_n t_0}. \tag{B.2}$$

where $H_0 |n\rangle = \zeta_n |n\rangle$, $\zeta_n > 0$ for $n > 0$, $\zeta_0 = 0$, and $w_n = \left(\zeta_n - i0^+\right)/\hbar$.

Since $\langle 0| p_{\ell\alpha} |0\rangle = 0$, the $n=0$ contributions are neglected and the $\chi$ matrix takes the form

$$\chi_{\ell\alpha\ell'\alpha'}(t_0) = \sum_{n=1}^{\infty} \langle 0| p_{\ell\alpha} |n\rangle \langle n| p_{\ell'\alpha'} |0\rangle e^{iw_n t_0}. \tag{B.3}$$

We seek the integral of this quantity from $t_0 = -\infty$ to $t_0 = 0$. The small dissipative term $i0^+$ accounts for dephasing from random effects; otherwise, the integral would not be defined since $e^{i\zeta_n t_0/\hbar}$ is a harmonic function. However, in a metal or semiconductor, the zero-point and thermal oscillations of the ion core lattice cause continuous dephasing of the electron system and $\chi_{\ell\alpha\ell'\alpha'}(t_0)$ will decay as $t_0 \to \infty$ with a rate that depends on the thermal conditions, and the properties of the lattice and its interaction with the electrons. With this, we obtain

$$\int_{-\infty}^{0} dt_0 \chi_{\ell\alpha\ell'\alpha'}(t_0) = -i \sum_{n \neq 0} \frac{\langle 0| p_{\ell\alpha} |n\rangle \langle n| p_{\ell'\alpha'} |0\rangle}{w_n}. \tag{B.4}$$

Note that for $\ell = \ell'$, $\alpha = \alpha'$, $i$ times this quantity is a positive real number. Furthermore, $i$ times this matrix (i.e., with indices $\ell\ell'\alpha\alpha'$, is Hermitian and hence has real eigenvalues).

If $|n\rangle$ $(n = 0,1,\cdots)$ is fermionic and is independent of $\underline{R}$, then the expectation values of (B.4) $\left(\langle 0| p_{\ell\alpha} |n\rangle \text{ and } \langle n| p_{\ell\alpha} |0\rangle\right)$ are independent of the particle label since the $|n\rangle$'s are

antisymmetric and therefore one can safely exchange labels inside the integral without affecting the final results. With this, $\tilde{\chi}_{\alpha\alpha'}$ takes the form

$$\tilde{\chi}_{\alpha\alpha'} = \frac{\hbar}{m^2} \sum_{n \neq 0} \frac{\langle 0|p_\alpha|n\rangle \langle n|p_{\alpha'}|0\rangle}{w_n}. \tag{B.5}$$